\documentclass[manuscript]{aastex}

\usepackage{float}

\shorttitle{}
\shortauthors{Sharykin and Kosovichev}

\begin{document}

\title{Investigation of Relationship Between High-Energy X-ray Sources and Photospheric and Helioseismic Impacts of X1.8 Solar Flare of October 23, 2012}

\author{I.N. Sharykin\altaffilmark{1,2}, A.G. Kosovichev\altaffilmark{3},
        V.M.~Sadykov\altaffilmark{3}, I.V.~Zimovets\altaffilmark{1}, I.I.~Myshyakov\altaffilmark{2}}

\affil{Space Research Institute of RAS, Moscow 117997, Russia}

\altaffiltext{1}{Space Research Institute (IKI) of the Russian Academy of Sciences}
\altaffiltext{2}{Institute of Solar-Terrestrial Research (ISTP) of the Russian Academy of Sciences, Siberian Branch}
\altaffiltext{3}{New Jersey Institute of Technology}

%%%%%%%%%%%%%%%%%%%%%%%%%%%%%%%%%%%%%%%%%%%%%%%%%%%%%%%%%%%%%%%%%%%%%%%%%%%%%%%%%%%%%%%%%%%%%%%%%%%%%%%%%%%%%

\begin{abstract}

The X-class solar flare of October 23, 2012, generated continuum photospheric emission and a strong helioseismic wave (``sunquake'') that points to an intensive energy release in the dense part of the solar atmosphere. We study properties of the energy release with high temporal and spatial resolutions, using photospheric data from the Helioseismic Magnetic Imager (HMI) onboard Solar Dynamics Observatory (SDO), and hard X-ray observations made by the Ramaty High-Energy Solar Spectroscopic Imager (RHESSI). For this analysis we use level-1 HMI data (filtergrams), obtained by scanning the Fe I line (6731~\AA) with the time cadence  of $\sim 3.6$ s and spatial resolution of $\sim 0.5^{\prime\prime}$ per pixel. It is found that the photospheric disturbances caused by the flare spatially coincide with the region of hard X-ray emission, but are delayed by $\lesssim 4$ seconds. This delay is consistent with predictions of the flare hydrodynamics RADYN models. However, the models fail to explain the magnitude of variations observed by the HMI. The data indicate that the photospheric impact and helioseismic wave might be caused by the electron energy flux substantially higher than that in the current flare radiative hydrodynamic models.

\end{abstract}

%%%%%%%%%%%%%%%%%%%%%%%%%%%%%%%%%%%%%%%%%%%%%%%%%%%%%%%%%%%%%%%%%%%%%%%%%%%%%%%%%%%%%%%%%%%%%%%%%%%%%%%%%%%%%
\section{INTRODUCTION}

Energy release during solar flares involves all layers of the solar atmosphere. Disturbances in the photosphere in the form of continuum emission are not frequent and usually observed in strongest flares. One of the most poorly understood phenomenon in the physics of solar flares is ``sunquakes'' initially detected by \cite{Kosovichev1998}. The basic properties and theories of sunquakes can be found in the reviews of \cite{Donea2011} and \cite{Kosovichev2014}. Generally, sunquake represents a helioseismic response, which is observed in Dopplergrams as concentric waves spreading out from an initial photospheric impact occurred during the impulsive phase of a solar flare. One scenario for initiation of helioseismic waves is a hydrodynamic impact caused by expansion of the chromospheric plasma heated by injection of nonthermal charged particles accelerated in the corona \citep[][]{Kosovichev1995}. The numerical hydrodynamic modeling of the beam-driven \textbf{thick-target} theory \citep[][]{Kostiuk1975, Livshits1981, Fisher1985, Kosovichev1986, Mariska1989, RubiodaCosta2014} predicts formation of a chromospheric shock that transfers the energy and momentum from the overheated chromospheric plasma to the colder and denser photosphere. This leads to compression and heating of the photosphere, and generation of helioseismic waves propagating into the deep solar interior.

Alternatively the plasma momentum can be transferred by other mechanisms, such as a sharp enhancement of the pressure gradient due to eruption of magnetic flux-rope \citep[e.g.][]{Zharkov2011,Zharkov2013} or by an impulse Lorentz force which can be stimulated by changing magnetic fields in the lower solar atmosphere \textbf{\citep{Hudson2008,Fisher2012,AlvaradoGomez2012,Burtseva2015,Russell2016}}. \cite{Sharykin2015a} and \cite{Sharykin2015b} discussed that rapid dissipation of electric currents in the low atmosphere could also explain sunquake initiation. However, it is possible that different sunquake events are caused by different mechanisms.

In this work we test the beam-driven theory of the photospheric impact and sunquake initiation. A large fraction of nonthermal electrons is thermalized in the chromosphere during their precipitation, and heats the chromosphere to very high temperatures. This process is accompanied by the HXR emission. After this, a shock wave (``chromospheric condensation'') is generated, and travels from the chromosphere to the photosphere, producing helioseismic waves. According to this model there should be a time delay between the HXR peak (indicating the maximum precipitation rate of the nonthermal electrons into the chromosphere) and the photospheric perturbation. Our task is to find this delay and its value from observations. Data with high temporal resolution are needed to find such delay. We use X-ray observations of RHESSI \citep{Lin2002}, and level-1 SDO/HMI data (filtergrams, \cite{Scherrer2012}). The HMI instrument scans the Fe I 6173~\AA~line in different polarizations to determine the line-of-sight (LOS) magnetic field, continuum intensity, Doppler velocity and the vector magnetic field in the photosphere. The intensity, Doppler velocity and LOS magnetic field maps are obtained with the time cadence of 45 sec, and do not have enough temporal resolution for measuring the time delay. The level-1 HMI filtergrams have the temporal resolution of $\approx 3.6$ s (for each of the two HMI cameras) which is comparable with the RHESSI time cadence of $\approx 4$ s, determined by the rotation period of spacecraft, needed for reconstruction of X-ray images. These data give us a chance for measuring the time delay.

For analysis we selected the X1.8 solar flare of October 23, 2012, started approximately at 03:13:00 UT. This flare was located in active region NOAA 1598 with heliographic coordinates S13E58. The flare generated strong helioseismic waves traveling from a large scale photospheric disturbance well seen in all HMI observables. So far, this is the strongest sunquake of Solar Cycle 24. Also, this flare produced an intensive photospheric emission source during the impulsive phase. To determine the time delay between the hard X-ray impulse and the photospheric impact we develop a special procedure for analysis of the HMI filtergrams. For comparison with the flare hydrodynamics simulations RADYN \citep{Allred2015} we use the NLTE radiative transfer code RH \citep{Pereira2015} to calculate the HMI line profile and its characteristics.

%%%%%%%%%%%%%%%%%%%%%%%%%%%%%%%%%%%%%%%%%%%%%%%%%%%%%%%%%%%%%%%%%%%%%%%%%%%%%%%%%%%%%%%%%%%%%%%%%%%%%%%%%%%%%%%%%%%%%%%
\section{VARIATIONS OF HARD X-RAY, PHOTOSPHERIC AND HELIOSEISMIC SIGNALS}

\begin{figure}[H]
\centering
\includegraphics[width=1.0\linewidth]{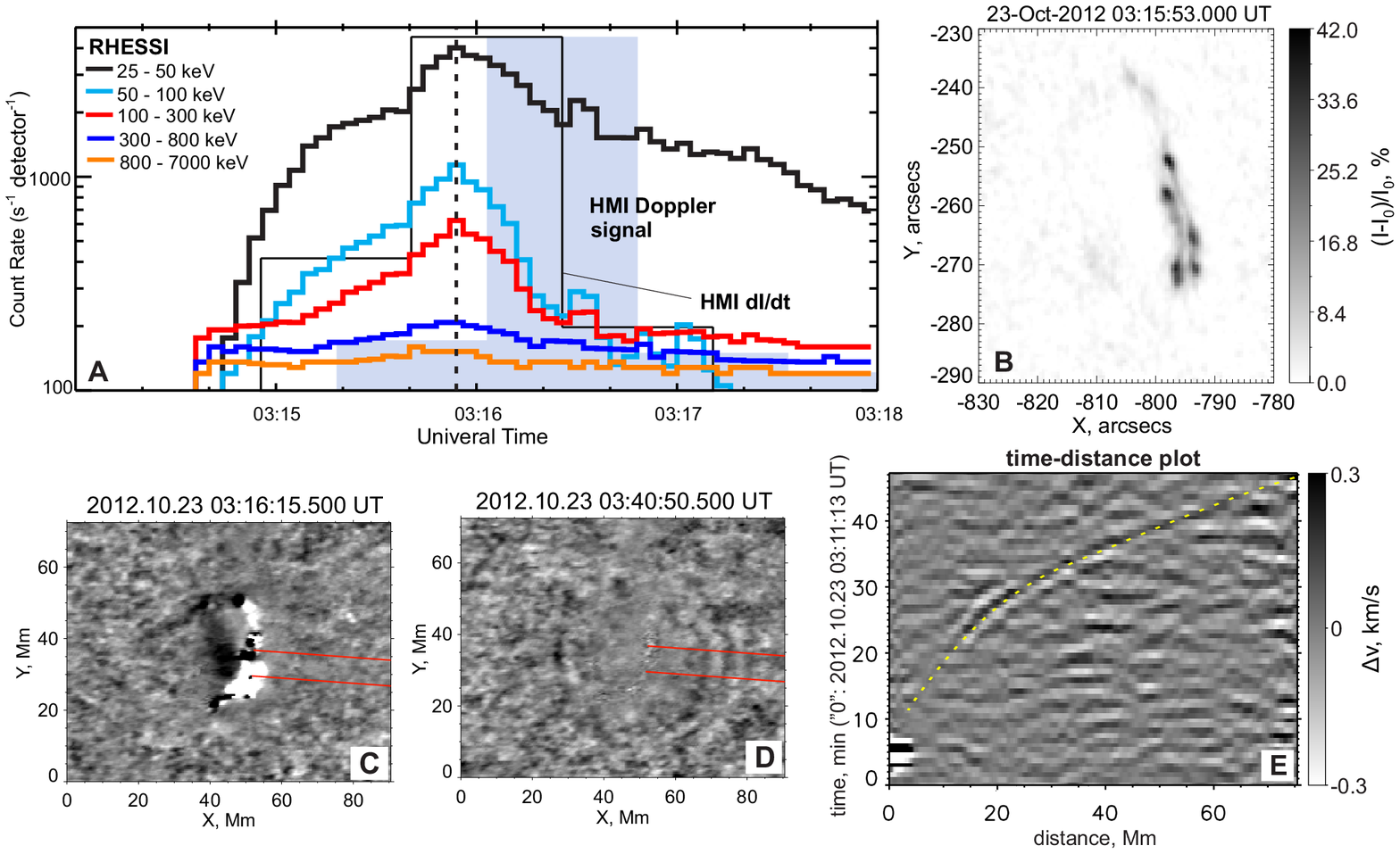}
\caption{Panel A shows the HXR RHESSI count rates in five energy ranges. The total positive Doppler speed (grey histogram) and time derivative of the HMI intensity $dI/dt$ (thin black line) are normalized to unity. Vertical dashed line corresponds to the HXR peak. Panel B presents the relative variation of the HMI continuum intensity during the HXR peak. Images in panels C and D are the time differences of Dopplergrams projected onto the Heliographic coordinates and filtered with a Gaussian frequency filter centered around 6 mHz showing the photospheric impact (C), and the helioseismic waves (D), 15 min after the impact. The sunquake time-distance diagram (panel~E) is calculated along the red lines shown in panel~D and compared with the ray-path theoretical prediction (dashed yellow line).}
\label{TP_SQ}
\end{figure}

Temporal variations of the hard X-ray and HMI observables: time derivative of relative intensity variations and the total positive Doppler speed signal in the flare region (grey histogram) are shown in Fig.\,\ref{TP_SQ}A. These HMI parameters are calculated as sums of all pixel's values in the region of significant enhancement of the continuum emission (Fig.\,\ref{TP_SQ}B). One can see that $dI/dt$ corresponds to the HXR peak. The Dopplergram signal enhancement in the flare region has a peak $\sim 30$ sec after the HXR peak. However, the 45-sec cadence of the HMI standard observables does not allow us to make a precise comparison with the 4-second RHESSI count rate.

Sunquake sources are usually associated with strong impulsive photospheric Doppler-shift signals and HXR sources \citep[e.g.][]{Kosovichev2006}. We present a time-difference Dopplergram (Fig.\,\ref{TP_SQ}C) calculated at the peak of the HMI continuum flare emission (Fig.\,\ref{TP_SQ}B) projected onto the heliographic coordinates and shown in the local Cartesian coordinates. The photospheric continuum emission enhancement as well as the Doppler-shift perturbation  during the HXR peak were observed as two ribbons $\sim 40^{\prime\prime}$ long and $\sim 5^{\prime\prime}$ apart. A significantly weaker continuum emission source was located about $\sim 15^{\prime\prime}$ east from the strongest source (Fig.\,\ref{TP_SQ}B). The entire photospheric emission pattern had a circular shape. One can notice that the sizes and shapes of the continuum emission source and the region of the enhanced Doppler-shifts are similar. The time profiles in Fig.\,\ref{TP_SQ}A reveal that the Doppler-shift perturbation is impulsive and precedes the peak of the total continuum emission. The circular shape of the photospheric emission sources is associated with a dome topology of the coronal magnetic field \citep{Yang2015}.

The sunquake event was initially revealed as a circular wave in the running difference of Dopplergrams remapped onto the heliographic coordinates (Fig.\,\ref{TP_SQ}D). This photospheric wave is a manifestation of acoustic waves traveling through the solar interior from a local flare-induced perturbation. To isolate the wave signal from convective noise we applied a Gaussian filter with a central frequency of 6 mHz and width of 2 mHz. Fig.\,\ref{TP_SQ}C illustrates the positions of the flare impact revealed as black-white regions corresponding to enhanced Doppler shift changes (positive or negative). The wave front is best visible about 15 minutes after the initial flare impact in the photosphere. The wave front amplitude is highly anisotropic (Fig.\,\ref{TP_SQ}D). The wave front traveling West had higher amplitude, than the wave front traveling East.

% We averaged the filtered Dopplergram signal between the lines to increase the signal-to-noise ratio.

The time-distance (TD) diagram for the strong helioseismic wave traveling West is shown in Fig.\,\ref{TP_SQ}E. Two solid red lines in panels C and D show the orientation of the image bands, along which we calculated the diagram. The bands are seven pixels wide, and oriented perpendicular to the wave front. We used these directions to calculate the TD diagram instead of circular averaging because the wave front was not completely circular due to the elongated source. Dashed curve in the TD~diagram represents the theoretical time-distance relation calculated in the ray approximation for a standard solar interior model of \cite{Christensen-Dalsgaard1993}. One can see that the position of the wave in the TD diagram is in accordance with the theoretical model.

\begin{figure}[H]
\centering
\includegraphics[width=1.0\linewidth]{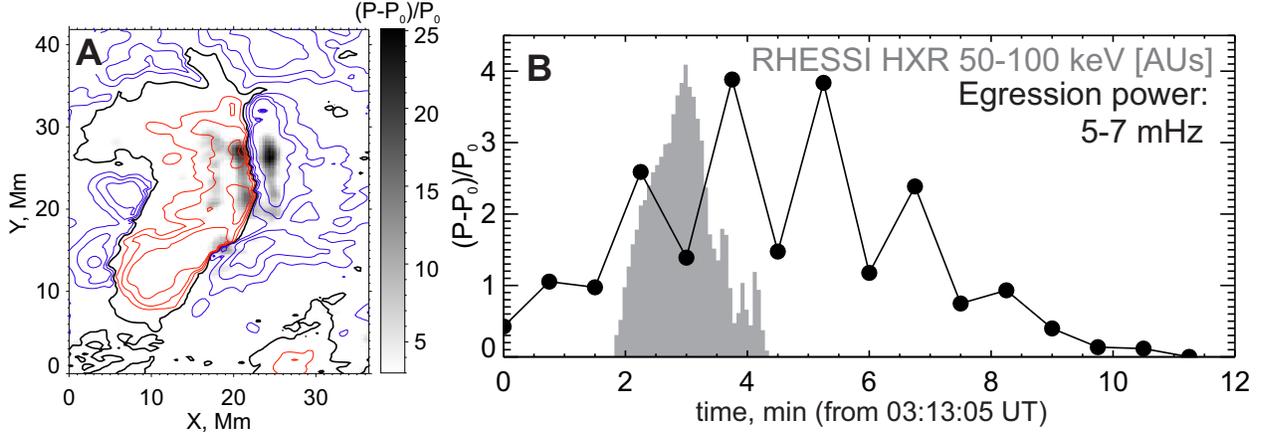}
\caption{A) Egression acoustic power map integrated over the flare impulsive phase around 6~mHz, relative to the egression power of the quiet Sun. Contour lines show magnetic field with levels of 0.5, 1 and 1.5 kG (red for $B_z>0$ and blue for $B_z<0$). Black curve shows the polarity inversion line (PIL). B) Temporal profile of the egression acoustic power (red) and the RHESSI count rate in the energy range of 50-100~keV.}
\label{Egress}
\end{figure}

To reconstruct the two-dimensional structure of the seismic source we used the helioseismic holography method \citep{Lindsey1997,Donea1999,Lindsey2000}. This approach uses a theoretical Green function of helioseismic waves to calculate the egression acoustic power corresponding to the observed velocity perturbations at various frequencies. The egression acoustic power map made in the frequency range of 5-7 mHz presented in Fig.\,\ref{Egress}A reveals two compact (less than 3~Mm in size) acoustic sources in the flare region located on both sides of the PIL (black curve in Fig.\,\ref{Egress}A). The PIL was determined using the HMI vector magnetogram reprojected onto the heliographic grid. The temporal profile (red curve in panel~B) of the egression acoustic power has a complex oscillating structure related to the helioseismic holography Fourier transform reconstruction procedure. However, one can see that enhancement of the egression power corresponds to the flare impulsive phase when RHESSI detected HXR emission (gray histogram in panel~B). Comparison of the sunquake source location with the X-ray signal and the photospheric impact is discussed in Section~4.

%%%%%%%%%%%%%%%%%%%%%%%%%%%%%%%%%%%%%%%%%%%%%%%%%%%%%%%%%%%%%%%%%%%%%%%%%%%%%%%%%%%%%%%%%%%%%%%%%%%%%%%%%%%%%%%%%%%%%%%%%%%%%%%%%%%%%%%%%
\section{ANALYSIS OF HMI FILTERGRAMS}

The two HMI cameras scan the Fe I line (6173~\AA), producing a series of images with the pixel size of 0.5$^{\prime\prime}$ for six wavelength positions across the line and in the near continuum, with a filter bandwidth (FWHM) $\sim 76$~m\AA~and the total wavelength tunning range $\sim 680$~m\AA~\citep{Couvidat2016}. Camera 1 takes images in linear polarization; and Camera 2 produces filtergrams in right and left circular polarizations. The filtegrams from both cameras are used to reconstruct the full Stokes profiles. For the line-of-sight magnetograms, Dopplergrams and continuum intensity (level-2 data), only Camera 2 is used. We use the filtergram data from both cameras to achieve a high temporal resolution in order to investigate variability of the photospheric flare emission sources. Previously, HMI filtergrams were used in the works of \cite{SaintHilaire2014} and \cite{MartinezOliveros2014}, for analysis of limb observations of flare loops. The time cadence of the level-1 data from each camera is 3.6 seconds, or 1.8 seconds for the sequence of filtergrams from both cameras. The RHESSI observations have the temporal resolution of 4 seconds determined by the spacecraft spinning period needed to obtain a full set of Fourier components used for image reconstruction.

\begin{figure}[H]
\centering
\includegraphics[width=1.0\linewidth]{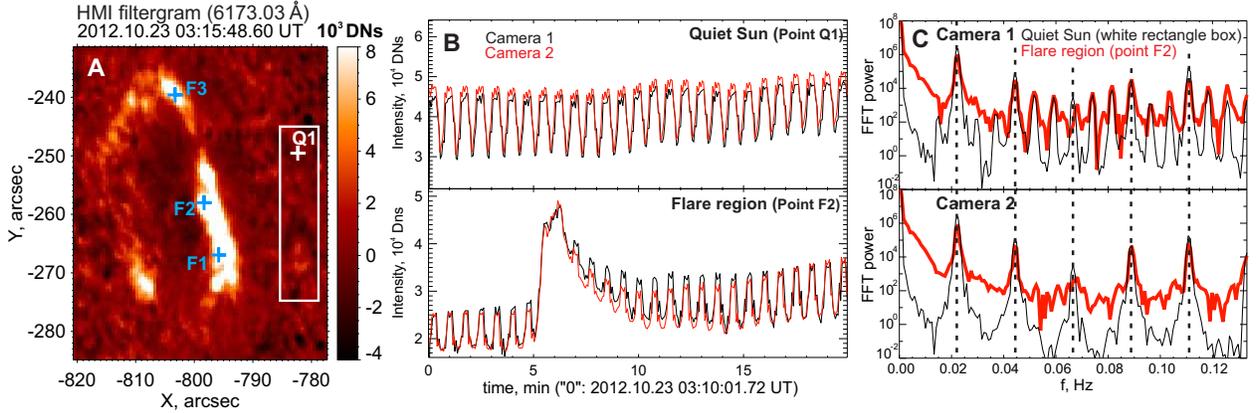}
\caption{Panel A shows an example of the flare HMI filtergram after subtraction of a preflare filtergram. Panel B presents the HMI filtergram lightcurves for a quite Sun region (point Q1 in panel C) in top subpanel, and for the flare region (point F2) in bottom subpanel, for Camera~1 (black) and Camera~2 (red). Panel~C shows the frequency power spectra for the quite Sun (black) and flare regions (red).}
\label{FFT}
\end{figure}

First, for the HMI data analysis we subtract preflare filtergrams from the filtergrams taken during the impulsive phase. This allows us to detect changes in the flare region. An example of such processed filtergram is shown in Fig.\,\ref{FFT}A. Then we compare the time profile of a quiet Sun region (point Q1) with the time profile of the flare region (point F2) shown in Fig.\,\ref{FFT}B. We present the time profile only for one pixel (Q1) because other quite Sun pixels show a similar behavior associated with the surface convection. The periodic variations of the line scans are due to the intensity changes across the line.

To remove the variations we apply a frequency filter. Figure~\,\ref{FFT}C shows the power spectra of the quiet Sun and flare regions for both cameras. One can see that the power spectrum for Camera~1 is more complex than the spectrum for Camera~2. There are 15 peaks above the $10^3$ threshold for Camera~1 and only 7 peaks for Camera~2. The largest peaks are harmonics of the 45-sec periodicity of the line scanning, and the Camera~1 data have an additional modulation corresponding to the 135-sec observing cycle. One can see that the modulation peaks for the flare and quiet Sun regions coincide with each other. To reduce the instrument modulation we apply Gaussian filters with the width of 10 mHz centered at the modulation peaks: 14~filters for Camera~1 and 6 filters for Camera~2.

\section{COMPARISON OF HMI FILTERGRAMS WITH RHESSI OBSERVATIONS AND RECONSTRUCTED SUNQUAKE SOURCE}

\begin{figure}[H]
\centering
\includegraphics[width=0.9\linewidth]{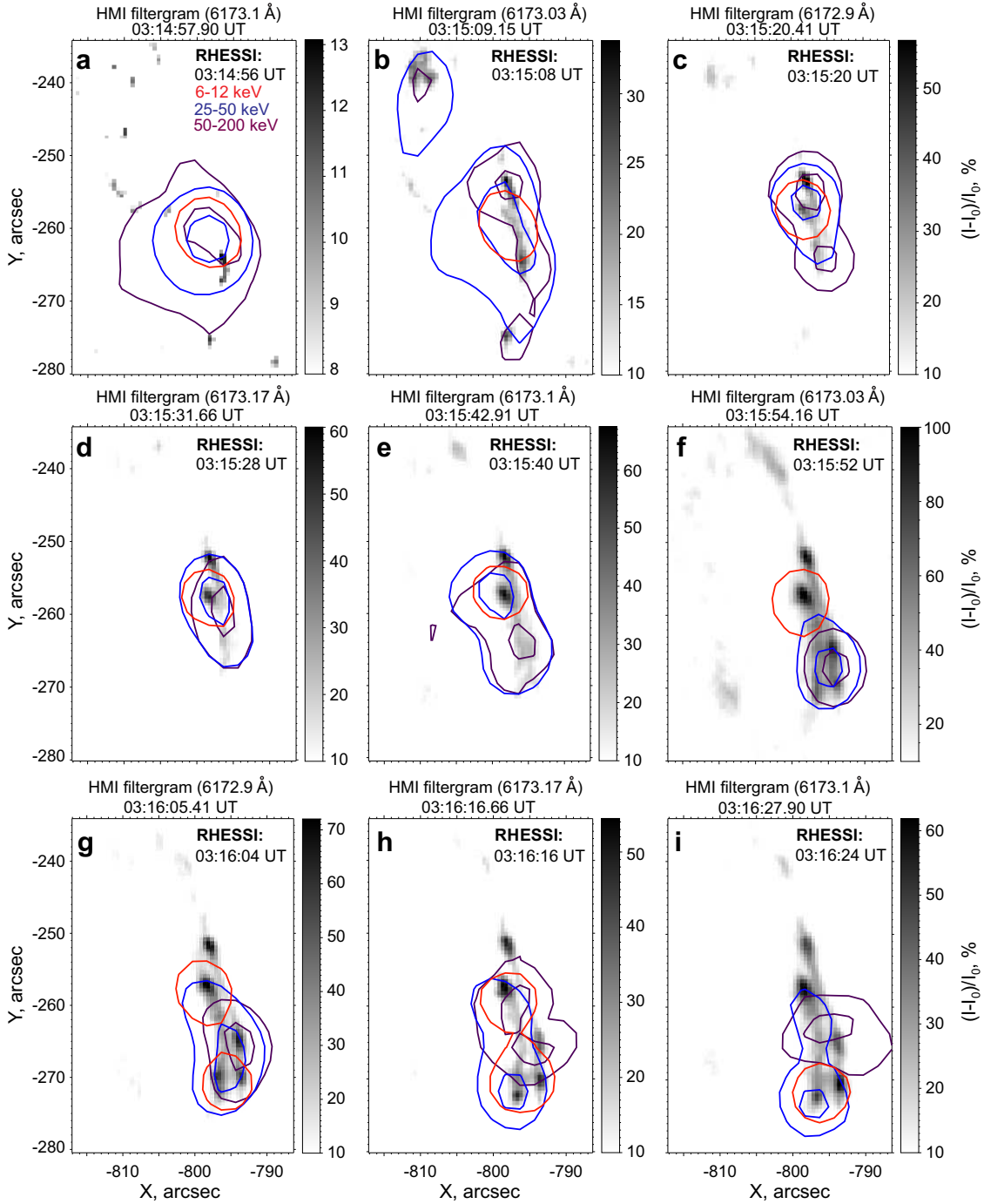}
\caption{Background images are the HMI filtergrams taken during the flare, after subtraction of a preflare filtergram. Black and blue contours show the corresponding RHESSI 25-50 and 50-200 keV contour maps for the 50, 70 and 90 \% levels relative to the maximum. Red contours show of the 6-12 keV X-ray sources at the 70~\% level.}
\label{filter}
\end{figure}

The filterd HMI filtergrams allow us to study the flare development with high temporal resolution. The initial photospheric flare perturbation was observed as small-scale weak photospheric brightenings (filtergram at 03:14:57 UT in Fig.\,\ref{filter}a). Then the flare emission is observed in four compact sources (03:15:09~UT, Fig.\,\ref{filter}b). Later we observe a strong emission enhancement in the form of large scale ribbons (03:15:54~UT, Fig.\,\ref{filter}f). The largest enhancement of the photospheric emission was approximately during the HXR peak and was about $\sim 100$\% above the quite Sun background.

In Fig.\,\ref{filter}, we compare the photospheric flare emission sources with the HXR sources from the RHESSI observations. The HXR images are synthesized in three energy bands: 6-12, 25-50 and 50-200 keV, using the CLEAN algorithm \citep{Hogbom1974,Hurford2002}. RHESSI Detectors 2,3,5,6 and 7 are used for the analysis. The high RHESSI count rate in this flare allows us to reconstruct the HXR images with the 4-second resolution and the pixel size of 2$^{\prime\prime}$. The HXR image field of view is $100^{\prime\prime}\times 100^{\prime\prime}$. The HXR sources were very dynamic during the flare impulsive phase. There were several HXR sources located in different parts of the photospheric flare ribbons. Generally, the 25-50 keV sources did not exactly coincide with the 50-200 keV sources. Sometimes positions of the HXR 25-50 keV sources correlate with the 6-12 keV emission sources (e.g. panels~h and~i). At the beginning of the impulsive phase (03:15:08 UT) the HXR sources nicely fit the photospheric enhanced intensity kernels (Fig.\,\ref{filter}b,\,c). During the maximum (approximately at 03:15:54 UT) the HXR emission source was relatively compact, had approximately circular shape with diameter of $\sim 10^{\prime\prime}$, and was located at the south end of the photospheric ribbon.

\begin{figure}[H]
\centering
\includegraphics[width=1.0\linewidth]{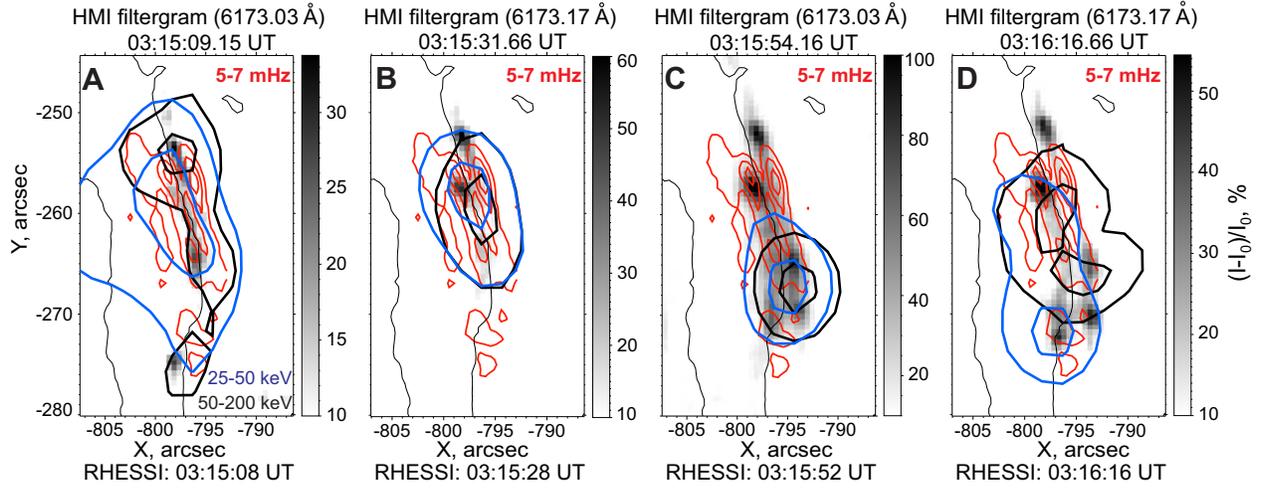}
\caption{Comparison of the HMI filtergrams (processed as described in Sec.~3) and RHESSI HXR images in the energy ranges of 25-50 (blue) and 50-200~keV (black) at four moments of time during the impulsive phase with the egression acoustic map (red contours with 20, 60 and 80 \% levels). The PIL is marked by black thin line.}
\label{EgRhFi}
\end{figure}

In Fig.\,\ref{EgRhFi}A-D we compare positions of the acoustic egression sources (red contours) derived from the helioseismic holography procedure, described in Sec.~2, with the HXR sources and the photospheric emission sources from the HMI high-cadence filtergram data obtained in Sec.~3. The acoustic source map is computed for the flare impulsive phase. One can see that the photospheric and HXR emission originated from the same region in the vicinity of the PIL, where the helioseismic waves originated. Two strong acoustic sources correspond to the two emission ribbons observed in the HMI filtergrams. Due to the relatively low spatial resolution and dynamic range of the RHESSI images we cannot make a more precise comparison of the HXR sources with the HMI observations. However, during the first half of the impulsive phase the strongest HXR and photospheric emissions coincide with the strongest sunquake source.

\begin{figure}[H]
\centering
\includegraphics[width=1.0\linewidth]{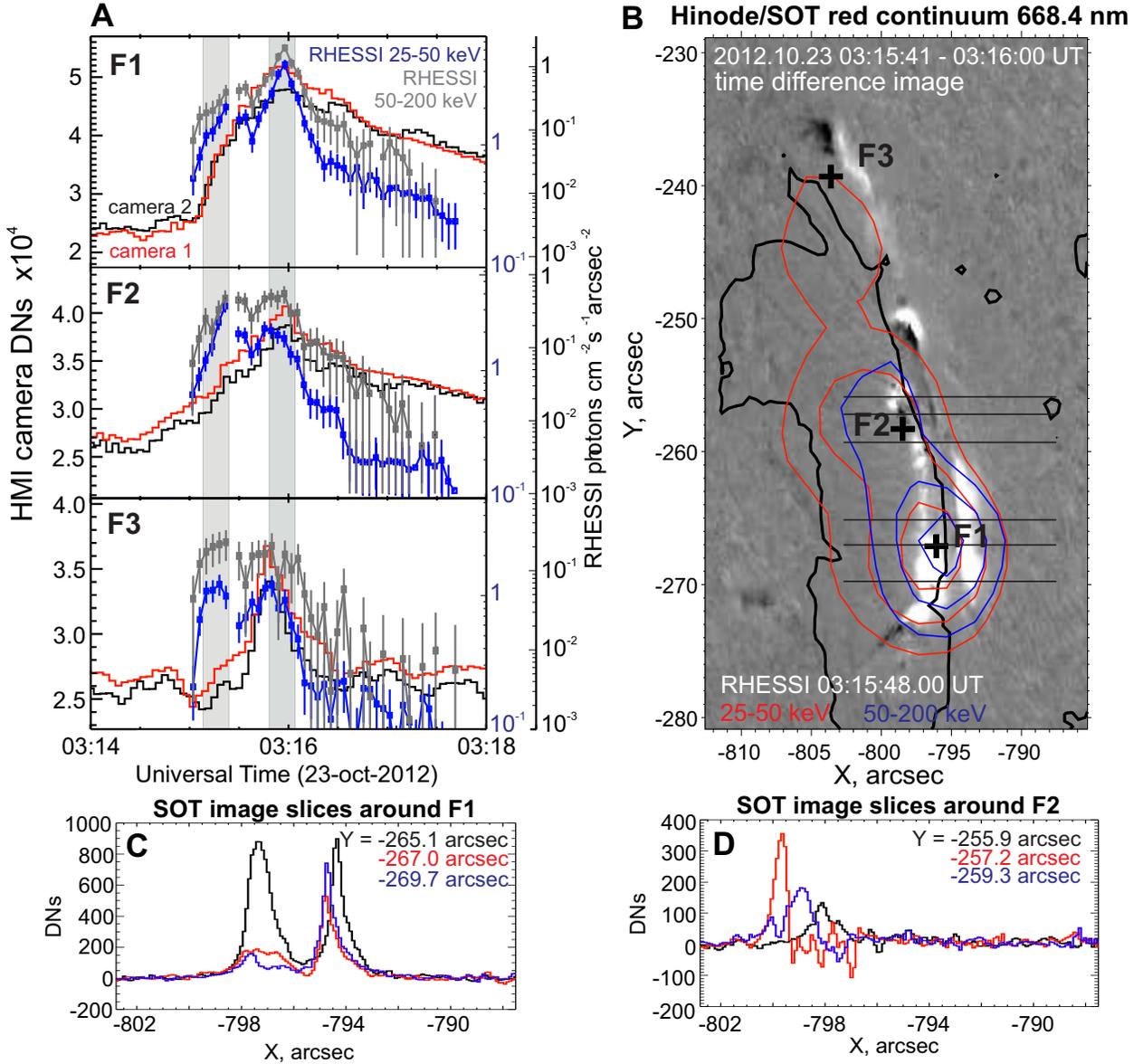}
\caption{Panel~A shows the filtered HMI filtergram lightcurves for three flare points (crosses marked as F1, F2, and F3 in panel~B) from camera 1 (red) and 2 (black). Grey and blue points with error bars show the total RHESSI fluxes in the energy ranges of 25-50 and 50-200 keV from the regions around selected points. Two gray vertical stripes correspond to the time ranges for the spectral analysis (see Fig.\,\ref{im_spec}). Panel~B shows the Hinode/SOT red continuum time difference image and the RHESSI contours (the same as in the Fig.\,\ref{filter}) in the energy ranges of 25-50 and 50-200~keV. Panels C and D show slices of SOT time difference image (panel B) in X direction for three Y positions around two points F1 (C) and F2 (D). These slices are shown by the horizontal black lines in panel~B.}
\label{delay}
\end{figure}

The demodulated HMI filtergram time profiles are shown in Fig.\,\ref{delay}A for three flare points located in different parts of the flare ribbons (see Fig.\,\ref{FFT}A). Points F1 and F2 are located in the sunquake source area. Point F2 corresponds to the strongest sunquake source (Fig.\,\ref{Egress}A). In Fig.\,\ref{delay}B the selected flare points are shown in the Hinode/SOT (Solar Optical Telescope) time-difference red continuum (6684~\AA) image. We calculated photon fluxes in the region-of-interest (ROI, $6^{\prime\prime}\times 6^{\prime\prime}$ square region) around the selected points. We show these lightcurves by blue and grey colors which correspond to the energy ranges of 25-50 and 50-200 keV. The errors are calculated as $\sim \sqrt{I}$ (where I is HXR image pixel values) assuming the Poisson statistics \citep{Bogachev2005}. One can notice that the HXR emission in two considered energy ranges has a two peak structure in all three regions. The amplitude of these peaks are comparable with each other in points F2 and F3. The second HXR peak and the photospheric intensity maximum occurred almost simultaneously. The time delay of the photospheric perturbations relative to the absolute maximum of the HXR lightcurves is less than 4 seconds for points F1 and F2. In point F3 the absolute maximum of the HXR lightcurve was 40 seconds earlier than the maximum of the photospheric intensity enhancement.

%It is evident that the photospheric emission was associated with the second HXR pulse. What is the reason for the second episode of nonthermal electrons injection to be an efficient agent of the photospheric disturbance generation?

To understand differences between the two HXR peaks we present the RHESSI imaging spectroscopy results in Fig.\,\ref{im_spec}. The X-ray spectra were calculated for two different time intervals marked by grey vertical stripes in Fig.\,\ref{delay}A. These time intervals correspond to the HXR peaks. The thermal part ($\lesssim 20$ keV) was fitted by a single-temperature bremsstrahlung emission spectrum. The nonthermal part ($\gtrsim 20$ keV) was approximated by a double-power law, in which one power index was fixed at 1.5. It corresponds to the X-ray emission below low-energy cutoff $E_{low}$ of nonthermal electrons. The low-energy cutoff is simulated by a break in the spectrum. For higher energies, spectral index $\gamma$ is a free fitting parameter. For numerical stability, $E_{low}$ is fixed and equal to 20 keV. Without fixing $E_{low}$ the fitting procedure leads to large errors of the fitting parameters. The fitting parameters are summarized in the Fig.\,\ref{im_spec} caption.

Point~F1 is characterized by the strongest change in the HXR hardness. Spectral index $\gamma$ changed from 3.7 to 2.9. Points F2 and F3 reveal minor changes in the hardness. The energy fluxes of nonthermal electrons were calculated following \cite{Syrovatskii1972}. These energy fluxes at the three flare points were 1.4$\times 10^{29}$, 1.7$\times 10^{29}$, and 0.8$\times 10^{29}$~erg/s, respectively, during the first HXR peak. The second peak was characterized by lower nonthermal energies: 1.1$\times 10^{29}$, 1.5$\times 10^{29}$, and 0.7$\times 10^{29}$~erg/s.

To estimate the electron precipitation area one can assume that it represents a circular region with a diameter equal to the photospheric ribbon width. From the HMI filtergrams one can determine the width of the photospheric ribbons as $\sim 2-4^{\prime\prime}\approx 1.5-3$~Mm. This estimate gives the precipitation area of $\sim2-7\times 10^{16}$~cm$^2$. The nonthermal electron energy flux per unit area is about $1.4-5.6\times 10^{12}$~erg s$^{-1}$cm$^{-2}$ for the total flux of $10^{29}$~erg/s. Using the Hinode/SOT time difference images with the pixel size of about~0.1$^{\prime\prime}$, one can deduce the area of particle precipitation more accurately. Figures~\ref{delay}C,\,D show the image slices along the X direction for three Y positions around points F1 (panel C) and F2 (panel D). The ribbon width (FWHM of intensity profile peaks) is in the range of $1-2^{\prime\prime}\approx 0.7-1.5$~Mm. In this case the energy density flux is in the range of $0.6-2.6\times 10^{13}$~erg~s$^{-1}$cm$^{-2}$. These estimates represent the upper limit of the electron energy density flux. The lower limit of the energy density flux can be also calculated using the SOT time difference images. The emission area above a threshold of 200~DNs is about $6.7\times 10^{16}$~cm$^2$. Thus, the low limit for the energy density flux is $\sim1.5\times 10^{12}$~erg s$^{-1}$cm$^{-2}$.

To demonstrate that the nonthermal electrons can generate the helioseismic waves we compare their energies with the energy of the helioseismic perturbation determined by integrating acoustic energy flux $c_s\rho\delta v^2/2$ in the egression source. Here, $\rho \approx 10^{-8}$ g~cm$^{-3}$ is the photospheric plasma density, $\delta v \approx 1-1.5$~km~s$^{-1}$ is the amplitude of plasma velocity perturbation determined from the acoustic egression maps, and $c_s \approx 10$~km~s$^{-1}$ is the photospheric sound speed. This gives the total helioseismic energy $E_{\rm seism}\approx 4\times 10^{28}$~erg which is approximately 1\% of the total energy of nonthermal electrons, $E_{\rm nonth}\approx 5\times 10^{30}$~erg, determined from the RHESSI X-ray spectra using formulas of \cite{Syrovatskii1972}. Thus, the nonthermal electrons have enough energy to generate the sunquake.

%  HXR FWHM contour has a circular shape. If we assume that area of the injection region has a circular shape.

%(counting pixels across the ribbon with intensity higher than 2000 DNs)

%However, according to Ca~II SOT/Hinode images from the work of \cite{Yang2015} the width of the ribbons can reach value of 1$^{\prime\prime}$ which results in electron energy density flux of order of $10^{13}$~erg s$^{-1}$cm$^{-2}$.

% One can see that the temperature did not experience significant changes when the emission measure significantly increased.

%%%%%%%%%%%%%%%%%%%%%%%%%%%%%%%%%%%%%%%%%%%%%%%%%%%%%%%%%%%%%%%%%%%%%%%%%%%%%%%%%%%%%%%%%%%%%%%%%%%%%%%%%%%%%%%%%%%%%%%%%%%%%%%%%%%%%%%%%%%
\section{DISCUSSION AND CONCLUSIONS}

We analyzed the helioseismic response (``sunquake'') of a X-class flare using HMI filtergrams and RHESSI X-ray data and found that the sources of the helioseismic waves were cospatial with the photospheric and HXR emissions during the flare impulsive phase. Using the HMI level-1 data (filtergrams) we determined positions of the photospheric intensity enhancements caused by the flare and extracted the time profiles with a 3.6-sec time cadence, and compared these with the high-cadence HXR data from RHESSI. The results showed that the time delay between the photospheric intensity and HXR peaks is less than 4 seconds.

\begin{figure}[H]
\centering
\includegraphics[width=1.0\linewidth]{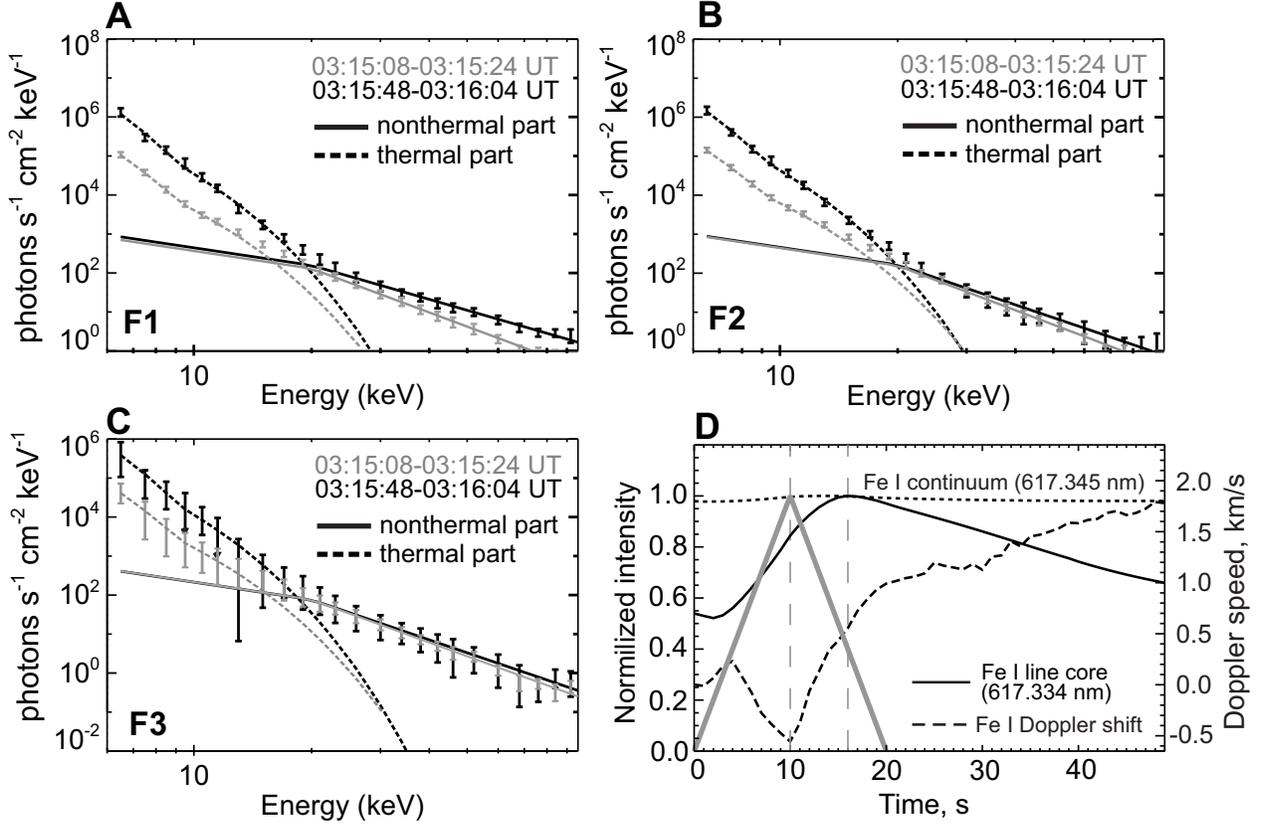}
\caption{Panels A-C show X-ray spectra for the selected flare points F1, F2 and F3 marked in Fig.\,\ref{delay}B. Grey and black curves within each panel correspond to two different time intervals marked by two gray vertical stripes in Fig.\,\ref{delay}-A. Point 1 (for two HXR peaks): $EM=$ $0.21\pm0.05$ and $3.57\pm0.85$ ($10^{49}$ cm$^{-3}$); $T=$ $2.4\pm0.2$ and $2.0\pm0.1$ (keV); $I_{10}=$ $371\pm50$ and $443\pm76$ (photons s$^{-1}$cm$^{-2}$keV$^{-1}$); $\gamma=$ $3.72\pm0.17$ and $2.89\pm0.18$. Point 2: $EM=$ $0.23\pm0.04$ and $3.87\pm0.80$ ($10^{49}$ cm$^{-3}$); $T=$ $2.64\pm0.16$ and $2.06\pm0.09$ (keV); $I_{10}$= $436\pm62$ and $461\pm103$ (photons s$^{-1}$cm$^{-2}$keV$^{-1}$); $\gamma=$ $3.79\pm0.17$ and $3.40\pm0.27$. Point 3: $EM=$ $0.08\pm0.07$ and $1.15\pm0.92$ ($10^{49}$ cm$^{-3}$); $T=$ $2.42\pm0.61$ and $2.00\pm0.36$ (keV); $I_{10}=$ $213\pm45$ and $214\pm86$ (photons s$^{-1}$cm$^{-2}$keV$^{-1}$); $\gamma=$ $3.63\pm0.23$ and $3.42\pm0.41$. Panel D presents results of the radiative hydrodynamic modeling from the RADYN database. Grey lines show the normalized nonthermal electron flux. Dotted and solid curves correspond to the normalized Fe I continuum (6173.45~\AA) and line core (6173.34~\AA) intensity. Dashed curve shows the Fe I Doppler velocity.}
\label{im_spec}
\end{figure}

To compare with theoretical predictions we use results of the radiative hydrodynamic flare modeling from the RADYN database \citep[http://www.fchroma.org, ][]{Allred2015}. The energy spectrum of nonthermal electrons is described by a power-law function with $\delta =4$, the peak energy density flux of $10^{11}$~erg~s$^{-1}$~cm$^{-2}$ (the highest available), and the low energy cutoff of 25~keV. The temporal profile of the nonthermal electron flux has a triangular shape and duration of 20 seconds. We consider the model with the largest energy density fluxes of nonthermal electrons. In the RADYN database, this model provides the strongest response in the low atmosphere. Using the plasma parameters and velocities from the RADYN model as an input for the radiative transfer code RH \citep{Pereira2015} we calculate the Fe I 6173~\AA~line observed by HMI. The theoretical time delay between the electron flux peak and HMI line response is consistent with the observations (Fig.\,\ref{im_spec}D). This shows that the radiative hydrodynamic model has a good potential for explaining the photospheric impacts. However, the magnitude of the predicted impact is much weaker in the model than in the observations.

In the model the Doppler velocity perturbation shows a weak ($\sim 0.5$~km/s) downflow which almost coincides with the heating function, followed by a gradual upflow. The amplitude of the plasma velocity in the photosphere in the model does not exceed 0.1 km/s. It is clear that the RADYN flare model fails to explain the HMI continuum emission, and does not provide the momentum transfer that is needed for the sunquake initiation. Our analysis shows that if the photospheric and helioseismic response is caused by the hydrodynamic response of the electron beam-heated atmosphere then the electron energy flux must be $10^{12}-10^{13}$~erg~s$^{-1}$~cm$^{-2}$, that is much higher than in the currently available flare radiative hydrodynamic models. Further development of flare radiative hydrodynamics models and other possible mechanisms of sunquake initiation is needed for better understanding of the observed photospheric and helioseismic effects.

This work was partially supported by NASA NESSF Fellowship Grant~NNX16AP05H, and NASA grant~NNX14AB68G.

%%%%%%%%%%%%%%%%%%%%%%%%%%%%%%%%%%%%%%%%%%%%%%%%%%%%%%%%%%%%%%%%%%%%%%%%%%%%%%%%%%%%%%%%%%%%%%%%%%%%%%%%%%%%%%%%%%%%%%%

\bibliographystyle{apj}
%\bibliography{bibliography}

\clearpage
%%%%%%%%%%%%%%%%%%%%%%%%%%%%%%%%%%%%%%%%%%%%%%%---------Figures-----------%%%%%%%%%%%%%%%%%%%%%%%%%%%%%%%%%%%%%%%%%%%%%

%----------------- time profiles (RHESSI) and Photospheric impact ---------------------------------

%----------------------------- Acoustic Holography ------------------------------------------------

%----------------------------- FFT analysis of filtergrams ----------------------------------------

%-------------------------------- RHESSI filtergrams ----------------------------------------------

%-------------------------------- Egree Map + RHESSI + filtergrams ----------------------------------------------

%----------------------------- Delays between HXRs and HMI lightcurves ----------------------------

%----------------------------- Imaging spectroscopy ----------------------------------------------

\clearpage
\end{document}